# Concerning the variability of beta-decay measurements

P.A. Sturrock[a,*], E. Fischbach[b], A. Parkhomov[c], J.D. Scargle[d], G. Steinitz[e]

[a] Center for Space Science and Astrophysics, and Kavli Institute for Particle Astrophysics and Cosmology, Stanford University, Stanford, CA 94305-4060, USA
[b] Department of Physics and Astronomy, Purdue University, West Lafayette, IN 47907, USA
[c] Institute for Time Nature Explorations, Lomonosov Moscow State University, Moscow, Russia
[d] NASA/Ames Research Center, MS 245-3, Moffett Field, CA 94035, USA
[e] Geological Survey of Israel, Jerusalem, 95501, Israel

*Corresponding author. Tel +1 6507231438; fax +1 6507234840.
Email address: sturrock@stanford.edu

ABSTRACT
Many experiments have been carried out to study the beta-decay rates of a variety of nuclides, and many – but not all – of these experiments yield evidence of variability of these rates. While there is as yet no accepted theory to explain patterns in the results, a number of conjectures have been proposed. We discuss three prominent conjectures (which are not mutually exclusive) – that variability of beta-decay rates may be due to (a) environmental influences, (b) solar neutrinos, and (c) cosmic neutrinos. We find evidence in support of each of these conjectures.

**1 . Introduction**

There are now several experiments that have been running over a sufficiently long period of time, producing enough data of high enough quality, to suggest that beta-decay rates of some nuclides are sometimes variable. One of the authors (AP), of the Lomonosov Moscow State University (LMSU), has performed experiments since 2000, exploring up to 7 different alpha and beta radio nuclides [1 – 3]. The measurements are performed around the clock with a periodicity of a few minutes. Detectors and power supplies are in thermostats. Temperature, atmospheric pressure, humidity, and background radiation around the installation are monitored continuously [4]. (See Figure 1.) It is notable that none of these experiments has yielded evidence for the variability of alpha decays, even though these experiments were otherwise identical to those that yielded evidence for the variability of beta decays.

Another author (GS), at the Geological Survey of Israel (GSI) in Jerusalem, has been running an experiment concerning radon for 8 years [5 - 11]. This experiment records 5 nuclear and 2 environmental measurements at 15-minute intervals, so producing (to date) a database with over 200,000 entries. This experiment also includes alpha measurements: unlike the AP experiments, the GS experiments show some alpha measurements to be variable. However, since the U238 decay chain (that leads to radon) involves both beta decays and alpha decays, it is not yet clear whether the variability in the alpha measurements is due to intrinsic variability of the alpha process, or whether it reflects the influence of variability of the beta process.

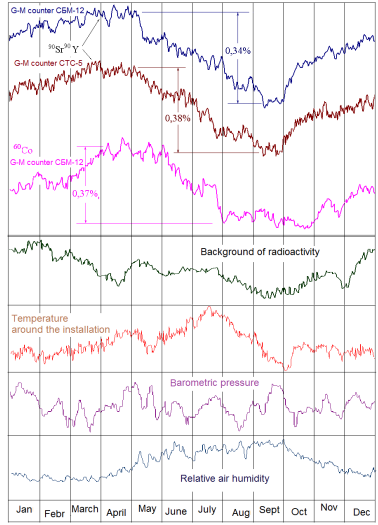

Figure 1. The average count rate of beta sources and major environmental parameters as functions of the annual period [4]. The averaging covers the results of the measurements obtained from 2000 to 2007.

It is important to note that patterns of variability are typically intermittent. The variability during one year may differ significantly from the variability during the preceding year. This intermittency may contribute to an apparent conflict between the above results of long-term experiments and the null results of some comparatively short-term experiments [12 - 14].

The overall problem of beta-decay variability remains in an early state with respect to theory. There are as yet no well-defined theories to be tested against the experimental data, but there are a few conjectures.

Rutherford proposed, as a result of his experiments, that nuclear decay rates intrinsically do not vary [15]. Siegert and his collaborators at the Physikalisch-Technische Bundesanstalt (PTB) found annual oscillations in the measured decay rates of Eu152 and Ra226, and suggested that these oscillations may not be intrinsic, but may rather be due to the effect of radon on the ionization chamber [16]. (We refer to this as the *PTB1* experiment.) More recently, Kossert and Nahle (also of PTB) have proposed that variations in beta-decay measurements are typically due to variations in environmental conditions [13,14,17]. We refer to this as the *Environmental Conjecture*.

Falkenberg, in a study of the beta decay of tritium, found evidence for an annual oscillation that appears to be related to the annually changing distance of the Earth from the Sun, and suggested that this effect may be due to solar neutrinos [18]. We refer to Falkenberg's suggestion as the *Solar-Neutrino Conjecture*.

It is interesting that Falkenberg draws attention to a speculation by Nikola Tesla (in 1932, before the concept of a neutrino was well established) that *radioactivity might be caused by small particles which are omnipresent and capable of passing any (non-radioactive) matter almost without leaving any traces* [18]. Falkenberg noted that although the peak in the oscillation (February) was close to the phase one would expect if the oscillations were due to the varying Earth-Sun distance, the amplitude in his experiment (0.37%) is only about one ninth of what would be expected (3.3%). This led Falkenberg to hypothesize that there may be a neutrino flux from other sources, the effect of which is approximately 8 times stronger than that from the Sun.

Together with Jenkins and other colleagues, EF has noted that measurements of the decay of Ra226 acquired at PTB [16] and measurements of the decay of Cl36 and Si32 acquired at the Brookhaven National Laboratory (BNL) [19] exhibit an annual oscillation similar to that expected of an Earth-Sun orbital influence on a solar neutrino flux [20,21]. This interest was stimulated by an apparent association between a great solar flare on December 13, 2006, and a simultaneous change in a beta-decay rate [22]. However similar flares have not led to similar changes in decay rates [23 - 24].

AP has pointed to the possibility that beta-decay rates may be influenced by cosmic neutrinos of very low energy instead of, or possibly in addition to, solar neutrinos [1,3,25]. This proposal seems to be compatible with the point raised by Falkenberg that, while decay rates may be influenced by solar neutrinos, there appears to be another factor that is several times more important. This other factor may be cosmic neutrinos. We refer to this suggestion as the *Cosmic-Neutrino Conjecture.*

It is important to note that these three conjectures do not form a "complete set": it is possible that some other mechanism or mechanisms may be responsible for the apparent variations in nuclear decay rates. For instance, nuclei may be affected by another form of radiation, and it is possible that some other particle (a boson) may couple neutrinos and radioactive nuclei. We should also note that these three conjectures are not mutually exclusive: it is possible that, for any one nuclide, the decay-rate may be influenced by two of the three factors or even by all three. Hence evidence in support of one of these conjectures (e.g. the Environmental Conjecture) does not constitute evidence against either of the others.

We now show that it is possible to find evidence in favor of each of these conjectures.

**2. The Environmental Conjecture.**
In analyzing any data set, it is essential to bear in mind the possibility that variability of the measurements may be due to the influence of environmental factors such as temperature, pressure, voltage, etc., on the measurement process. (It is of course also possible that environmental factors may actually influence the decay process, a question that can be addressed only after one has obtained a complete understanding of the influence of the environment on the measurement process.)

Since beta-decay experiments require the measurement of fluctuations of order one-tenth or one-hundredth of a percent, it is essential to take great care to control environmental influences and if possible to use detectors that are resistant to these influences. It is also important to monitor environmental parameters. For instance, AP's experiments provide for continuous measurements of atmospheric temperature, pressure, humidity and radiation background [4]. (See Figure 1.)

We may ask whether it may be possible to determine that measurements have in fact been influenced by environmental factors. One possibility would be to run two identical experiments, one of which is subject to environmental influences and the other not. The former experiment may be run above ground and the latter deep below ground, controlled for temperature and pressure and shielded from radiation (such as might be produced by cosmic rays). If the former shows an annual oscillation but the latter does not, it would be reasonable to conclude that the positive result was due to environmental factors. (However, if neither shows an annual oscillation, the experiment would be inconclusive.)

Another possibility is to run two identical experiments, one in the northern hemisphere and the other in the southern hemisphere. If they exhibit annual oscillations that are in anti-phase, that would tend to support the environmental conjecture.

One may also examine the results of experiments that cycle through two or more nuclides, examining them all with the same equipment. If the dominant influence is environmental, the measurements should be similar. In particular, they should all show similar annual oscillations. This was in fact the philosophy behind the original BNL experiment of Alburger et al. [19] who measured the decay rates of Si32 and Cl36 in the same apparatus over a 4-year period. If the annual variations that they separately measured in these two isotopes were the result of an annual variation in the sensitivity of their common detector, then this oscillation should have cancelled when the Si32/Cl36 ratio was determined. However, the oscillation is even more prominent in the ratio, which we see as a strong argument against the *Environmental Conjecture* with respect to this experiment.

In this context, is interesting to review the results of an analysis of the PTB1 experiment [26,27] and of the results of an analysis [28] of a more recent experiment carried out at PTB, which we refer to as *PTB2* [29]. Each experiment examined several nuclides in a cyclic manner with the same measuring equipment. We list the amplitude and phase of maximum for each nuclide in Tables 1 and 2, and we display these data in Figures 2 and 3. We see that, for PTB1, the amplitudes and phases are remarkably consistent. For PTB2, however, the amplitudes are much smaller and much more variable, as also are the phases. This strongly suggests that the PTB1 measurements were dominated by environmental processes (as suggested by Schrader [29]), but the same is not true of the PTB2 measurements. The points in Figure 3 resemble a scatter diagram, which is consistent with the amplitudes and phases being intrinsic to the nuclides, not being due to environmental or experimental influences. On examining the powers listed in Table 2, there appears to be significant evidence for intrinsic annual oscillations in data acquired for Ag108, Eu152, and Ra226. These results suggest that different nuclides respond very differently to whatever is responsible for the annual modulations.

Table 1. For the PTB1 dataset, and for each nuclide and for the annual oscillation, the power, amplitude, and phase of maximum of the normalized current.

| Nuclide | Power | Amplitude | Phase |
|---------|-------|-----------|-------|
|         |       |           |       |
| Ag108   | 125   | 0.000864  | 0.0789 |
| Ba133   | 60    | 0.000680  | 0.0808 |
| Eu152   | 127   | 0.000827  | 0.0713 |
| Eu154   | 122   | 0.000813  | 0.0684 |
| Kr85    | 113   | 0.000741  | 0.0616 |
| Ra226   | 122   | 0.000852  | 0.0707 |
| Sr90    | 115   | 0.000877  | 0.0832 |
|         |       |           |       |
|         | mean  | 0.000808  | 0.0735 |
|         | stdev | 0.000072  | 0.0077 |

Table 2. For the PTB2 dataset, and for each nuclide and for the annual oscillation, the power, amplitude, and phase of maximum of the normalized current.

| Nuclide | Power | Amplitude | Phase |
|---------|-------|-----------|-------|
|         |       |           |       |
| Ag108   | 24.83 | 0.000155  | 0.093305 |
| Ba133   | 3.33  | 0.000118  | 0.382917 |
| Cs137   | 12.59 | 0.000092  | 0.485868 |
| Eu152   | 35.19 | 0.000244  | 0.528080 |
| Eu154   | 1.81  | 0.000042  | 0.219202 |
| Kr85    | 4.17  | 0.000081  | 0.665972 |
| Ra226   | 68.31 | 0.000198  | 0.671191 |
| Sr90    | 7.29  | 0.000103  | 0.241110 |
|         |       |           |       |
| mean    | 19.69 | 0.000129  | 0.410955 |
| stdev   | 22.87 | 0.000066  | 0.213605 |

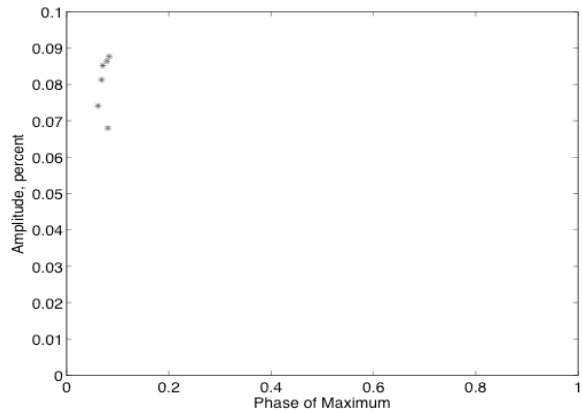

Figure 2. Display of the amplitude and phase of maximum of the annual oscillation of seven nuclides analyzed in the PTB1 experiment (two points overlap).

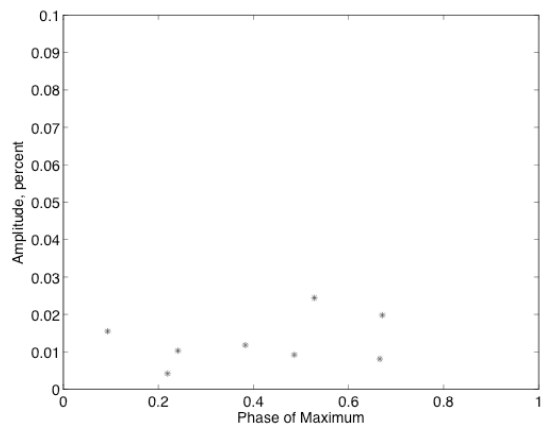

Figure 3. Display of the amplitude and phase of maximum of the annual oscillation of eight nuclides analyzed in the PTB2 experiment.

## 3. The Solar-Neutrino Conjecture.

An annual oscillation may be due to a solar influence, but it may also be due to an environmental influence. Hence, to test a solar conjecture, one should look for other evidence. One obvious possibility is to search for oscillations that may be associated with solar rotation. If the influence is due to neutrinos, modulation of the neutrino flux may in principle occur anywhere inside the Sun. Theoretically, the neutrino flux could be influenced by either the MSW (Mikheyev, Smirnov, Wolfenstein) effect [30,31] that is determined by the density structure, or the RSFP (Resonant Spin Flavor Precession) effect [32] that depends on both the density and the magnetic field.

The Sun is sufficiently stable that the MSW effect would not lead to any detectable time variation. On the other hand, the solar magnetic field, as it is observed at the photosphere, is highly asymmetric and highly variable. The same is likely to be true throughout the convection zone, and may also be true in the radiative zone. Hence if the RSFP process is operative in a region where the magnetic field is sufficiently strong and inhomogeneous, we may expect that the solar neutrino flux may exhibit modulation in a band of frequencies appropriate to the Sun's internal rotation. As determined by helioseismology, the equatorial sidereal rotation rate is in the range $13.5 – 15.0$ year$^{-1}$, which converts to a synodic rate (as seen from Earth) of $12.5 – 14.0$ year$^{-1}$ [33]. However, the rotation rate in the deep interior is uncertain and there are indications from analyses of Super-Kamiokande measurements that some part of the solar interior may rotate as slowly as $10.4$ year$^{-1}$ (sidereal) or $9.4$ year$^{-1}$ (synodic) [34]. Hence a reasonable search band for modulation of beta-decay rates would be $9 – 14$ year$^{-1}$, corresponding to periods in the range $26 – 41$ days. It is significant that there is no obvious environmental influence with a period in this range.

There is in fact evidence of modulations of beta-decay rates in this band. AP has found evidence of oscillations with periods of order 27 days [2]. PS and colleagues have found similar evidence [35], as have Javorsek et al. [36]. These oscillations tend to be transient, and are therefore best investigated in terms of spectrograms rather than periodograms. Figures 4 and 5 show spectrograms formed from BNL Cl36 and Si32 data, and Figure 6 shows for comparison a spectrogram formed from Super-Kamiokande data [37]. They all show a common feature at about $12.5$ year$^{-1}$, compatible with the influence of rotation in the radiative zone.

The Sun exhibits many kinds of oscillations, including one referred to as *r-mode oscillations* (which are referred to as *Rossby waves* by geophysicists) [38]. The frequencies of these oscillations are determined by *l* and *m* (two of the three spherical harmonic indices) and the local sidereal rotation rate. Analyses of two sets of solar diameter measurements have revealed evidence of several r-mode oscillations that have their origin where the sidereal rotation rate is $12.08$ year$^{-1}$ [39,40]. We have found evidence of similar oscillations in beta-decay data acquired by AP [41].

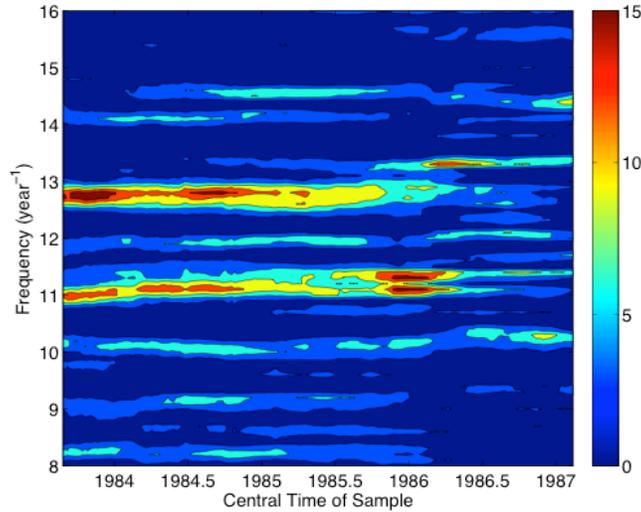

Figure 4. Spectrogram formed from BNL Cl36 data, showing a transient oscillation with frequency close to 12.5 year$^{-1}$. The colorbar shows the power of the oscillation.

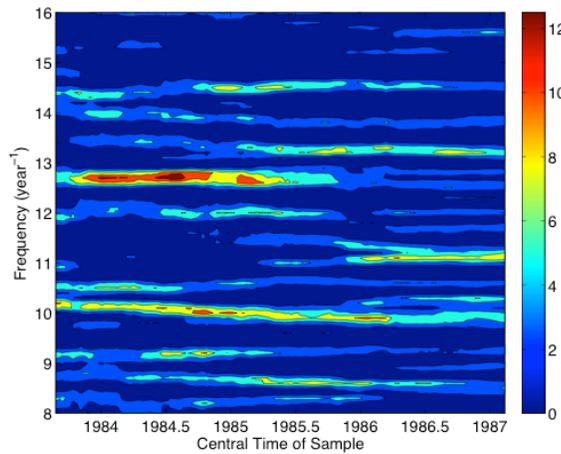

Figure 5. Spectrogram formed from BNL Si32 data, showing a transient oscillation with frequency close to 12.5 year$^{-1}$.

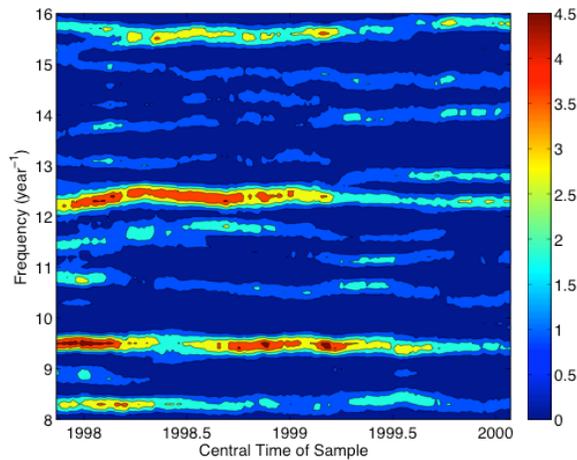

Figure 6. Spectrogram formed from Super-Kamiokande solar neutrino data, showing a transient oscillation with frequency close to 12.5 year$^{-1}$.

## 4. The Cosmic-Neutrino Conjecture.

This conjecture concerns the possibility that fluctuations in beta-decay rates may be due in part to the influence of cosmic (relic) neutrinos, which we expect to have low (non-relativistic) speeds. According to a theory developed by AP, the influence of slow neutrinos on the decay rate of a nuclide is expected to be sensitive to the relative velocity [25]. An attractive feature of the AP theory is that the fractional fluctuation in the count rate is independent of the decay rate, which is consistent with experimental results. The annual oscillations in decay rates may be a result of the annual variation of the velocity of the Earth relative to the cosmic-neutrino background, due to the Earth's orbital motion.

For the investigation of this conjecture, we find the most helpful dataset to be one compiled by Steinitz et al. [6], as analyzed in [11]. This article analyses 29,000 measurements of gamma radiation associated with the decay of radon in a sealed container at GSI between 28 January 2007 and 10 May 2010. These measurements were found to exhibit strong variations in both time of year and time of day. Time-series analysis revealed a number of periodicities, including one at 11.2 year$^{-1}$ and one at 12.5 year$^{-1}$, which are in close agreement with periodicities evident in the spectrogram formed from BNL Cl36 data (see Figure 4). Hence the analysis of GSI data supports the Solar-Neutrino Conjecture. However, the analysis also revealed an extremely strong annual oscillation.

Formatting measurements into one-hour intervals, it was possible to separately examine daytime and nighttime data. Figures 7 and 8 show displays of the power of an oscillation as a function of both frequency and time of day for the frequency bands 0 - 8 year$^{-1}$ and 8 – 16 year$^{-1}$, respectively. We see that the annual oscillation is primarily a daytime feature, centered on noon, whereas the "solar" oscillations (in the band 11 to 12.5 year$^{-1}$) are primarily nighttime features, centered on midnight.

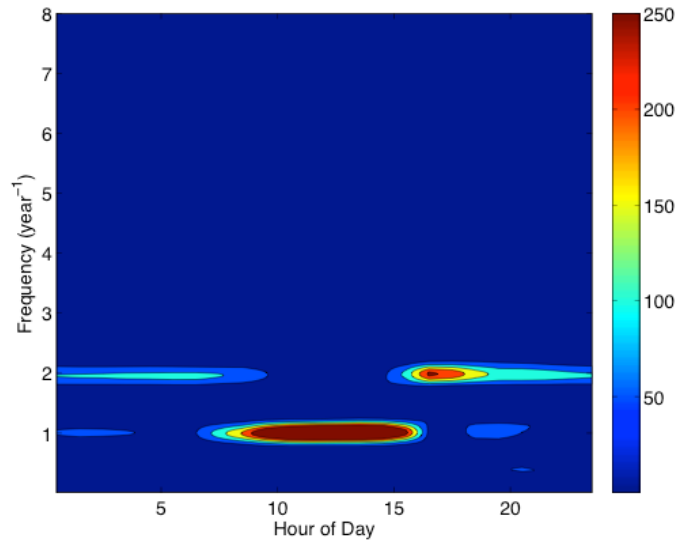

Figure 7. Spectrogram formed from GSI radon measurements, showing that an annual oscillation is evident in daytime measurements.

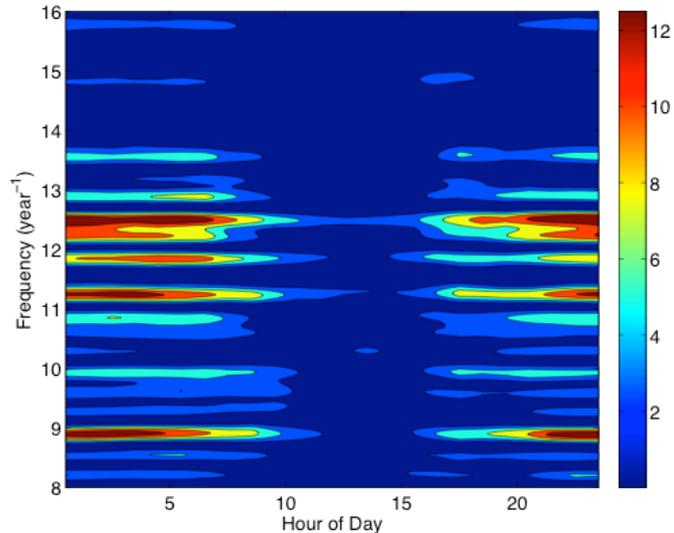

Figure 8. Spectrogram formed from GSI radon measurements, showing that an oscillation with frequency close to 12.5 year$^{-1}$ is evident in nighttime measurements.

The key to understanding this curious pattern is to examine Figure 21 of ref. [11], which shows the experimental layout. Radon (Ra222) has its origin in the decay of uranium (U238) in a layer of ground phosphorite at the bottom of a closed container. A gamma-ray sensor, located in the upper part of the tank, is contained in a lead pipe which has a perforated lead plate at its lower end. The plate serves to reduce the direct gamma radiation from the phosphorite, while allowing gas to enter the gamma-detector chamber.

The crucial point is that there is a *directional relationship* between the source of the gamma radiation (radon in the cylinder) and the gamma detector (located at the top of the cylinder). Hence the detector is more sensitive to photons traveling vertically upwards, and less sensitive to photons traveling vertically downwards. We can therefore understand the pattern shown in Figures 4 and 5 on the basis of the following two hypotheses:

(a) Decays are stimulated by neutrinos, and
(b) Photons resulting from a stimulated decay tend to travel preferentially in a direction parallel to the direction of the incoming neutrino.

According to these hypotheses, the gamma detector is preferentially responsive to the influence of neutrinos that are traveling vertically upwards. Neutrinos from the Sun will be traveling vertically upwards at midnight (the neutrinos having traveled through the Earth). This offers an explanation of the fact that oscillations associated with solar rotation show up in the nighttime measurements, not in the daytime measurements.

According to this scenario, the daytime measurements will be sensitive to neutrinos that are traveling *towards* the Sun. These can only be cosmic neutrinos. The further inference is that cosmic neutrinos in the solar system are traveling preferentially towards the Sun – presumably in response to the solar gravitational field. The outgoing neutrinos may tend to be isotropized by the gravitational encounter.

Further indications for directionality in the pattern of gamma radiation from the radon system are demonstrated in other experimental examinations [6, 10]

**5. Discussion.**

Of the various patterns that have emerged from analysis of the experimental results, the most significant appear to be the various indications of oscillations attributable to solar rotation. Figure 6 shows a spectrogram formed from Super-Kamiokande 5-day data. It shows evidence of an oscillation with frequency 12.5 year$^{-1}$, which is compatible with the influence of rotation in the solar radiative zone, as determined from helioseismology [33]. We find similar features in several datasets formed from decay experiments. AP found oscillations with a period of about one month (frequency 12 year$^{-1}$) in measurements of the decay rates of Cs137, Co60 and Si32 [3]. Spectrograms formed from BNL measurements of the decay of Cl36 and Si32, shown in Figures 4 and 5, respectively, show evidence of an oscillation with frequency 12.5 year$^{-1}$. Figure 8 is particularly interesting, since it shows evidence for an oscillation with frequency 12.5 year$^{-1}$ (and also nearby oscillations at 11.2 year$^{-1}$ and 11.9 year$^{-1}$), but these oscillations are evident only in measurements made near midnight which, as we have noted, is indicative of a solar stimulus. To the best of our knowledge, the only conceivable environmental influence with a similar frequency is that of the Moon, but a lunar source would not be compatible with the directionality evident in GSI measurements (evident in Figure 8). We are left with what appears to be strong evidence for a solar influence on beta-decay rates, supportive of the solar-neutrino conjecture.

On the other hand, many experiments – such as [18 - 23] - have yielded evidence of an annual oscillation. Figure 7, derived from GSI radon experiments [11], presents strong evidence for such an oscillation, most evident in daytime data, which is suggestive of a cosmic source rather than a solar source. This result is therefore supportive of the cosmic-neutrino conjecture.

Since beta-decay variability appears to be intermittent and to be different for different nuclides, and since different measurement techniques give different results, then with the goal of reviewing current conjectures, there seems to be a need for a new generation of experiments, for which we advance the following suggestions:

(a) Strict control of environmental conditions is essential. In particular, experiments should guard against the influence of radon, since radon decay appears itself to be variable.
(b) It is highly advantageous to examine several nuclides in the same experiment - either simultaneously, cyclically or on a randomized schedule. (As an example, the BNL experiment involved 20 measurements per day of Cl36 decay and 20 interleaved measurements per day of Si32 decay.)
(c) Tests to determine the dependence of measured decay rates on conditions in the laboratory, such as temperature, atmospheric pressure, humidity, etc. These tests may include the examination of environmental parameters not previously suspected as being important. The analysis of previously obtained data sequences to detect any dependence of measured decay rate on environmental parameters could advisedly be followed by sequences of measurements in which the environmental parameters in the laboratory are intentionally varied in a controlled way over ranges comparable with known annual variations.
(d) In order to assess the role of environmental influences, it would be helpful to include in the experiment a strictly alpha-decaying nuclide and a pulse generator configured to generate a count rate similar to that of the beta-decay specimens.

(e) With the goal of distinguishing spontaneous and induced beta decays, it will be essential to examine the variability of measurements as a function of the beta-particle energy, paying special attention to variability at and near the endpoint energy.
(f) There is evidence that the decay process is anisotropic. It is therefore desirable that there be an array of detectors to test for anisotropy and, if anisotropy is confirmed, to determine the relevant polar diagram as a function of polarization.
(g) Since any influence on beta decays would show up most clearly near the endpoint of the beta spectrum, where the electron energy is nearest its maximum value [21,42,43], one should preferentially use detectors that are most sensitive to high energies or, even better, determine the energy of each electron resulting from beta decay.
(h) It would be interesting to study the correlation function formed from the results of two identical experiments with varying spatial separation, so as to obtain spatial information about any particle or radiation field that may influence decay rates.
(i) If future experiments support the conjecture that a particle or radiation field influences beta decays, one could conceive of experiments designed to detect any macroscopic force or torque that the particles or field may exert on a nuclide [37].

Finally, we note that experiments designed to measure the mass $m_\nu$ of the electron antineutrino by studying the endpoint of the tritium decay energy spectrum may be candidates for detecting deviations from the expected electron spectrum due to neutrino-induced decays. It is interesting to note that many such experiments find an anomalous value for $m_\nu^2$ [44]. As noted in Ref. [21], this may already indicate the presence of neutrino-induced decays.